# Detection of Islands and Droplets on Smectic Films Using Machine Learning

Eric Hedlund, Keith Hedlund, Adam Green, Ravin Chowdhury,
Cheol S. Park, Joseph E. Maclennan, and Noel A. Clark

*Department of Physics and the Soft Materials Research Center,*

*University of Colorado, Boulder, Colorado 80309, USA*

Machine learning techniques have been developed to identify inclusions on the surface of freely suspended smectic liquid crystal films imaged by reflected light microscopy. The experimental images are preprocessed using Canny edge detection and then passed to a radial kernel support vector machine (SVM) trained to recognize circular islands and droplets. The SVM is able to identify these objects of interest with an accuracy that far exceeds that of conventional tracking software, especially when the background image is non-uniform or when the target features are in close proximity to one another. This method could be applied to tracking objects in a variety of visually inhomogeneous biological and soft matter environments, in order to study growth dynamics, the development of spatial order, and hydrodynamic behavior.

## I. INTRODUCTION

Since their discovery in the 1970s [1], ultrathin, films of fluid smectic liquid crystal freely suspended in air have been a continuous wellspring of new liquid crystal physics, providing a unique experimental platform for the study of phase behavior, fluctuations, elasto-hydrodynamics, and interfacial effects in 2D fluids [2]. The smectic fluid layering on the one hand enables such films to be drawn, and on the other hand stabilizes their structure, enabling films of a variety of thicknesses to be prepared, from hundreds of layers down to a single molecular layer [3]. While it is possible to prepare films of uniform thickness, in many cases the films of interest are inhomogeneous, with edge dislocations in the layering ("layer steps"). In addition, the films may contain inclusions such as pancake-like smectic islands and isotropic droplets, either occurring naturally or created intentionally during the process of preparing or heating the film [4,5]. The motion of such inclusions has been used to investigate the fundamental hydrodynamic properties of smectic films [6,7,8,9,10,11,12], and their spatial organization studied in order to better understand their interactions [5,13,14]. In tilted phase films, the study of the dynamics of topological defects in the director field has also proven to be a rich trove of interesting fundamental physics [15,16,17,18,19].

The OASIS experiments, carried out on the International Space Station in 2017, yielded extensive observations of the coarsening, coalescence, and thermomigration of emulsions of islands and droplets on the surface of spherical smectic bubbles in microgravity [20, 21, 22,23]. Analysis of the images from this mission has proven to be scientifically interesting but non-trivial. The principal challenges being to distinguish features of interest on the entire surface of the transparent bubble and to classify such features correctly even if they are partly or temporarily obscured, or degraded in appearance by imaging artefacts.

Tracking the motion of objects of interest in a series of experimental images typically involves the following steps: (1) preprocessing the images in order to make the objects easier to identify; (2) determining the locations of the target objects in successive images; and (3) linking these locations in order to determine their motion over time. One of the most widely used tracking applications in the broader soft matter community is the Python software TrackPy [24], which implements techniques for identifying and tracking features in experimental images based on algorithms originally developed for colloidal particle tracking in IDL by Crocker and Grier [25,26]. In this method, global thresholding is typically used to obtain high-contrast binary images and then target objects with a predetermined range of sizes are identified and tracked. The TrackPy software is easy to learn and run, is well documented, and can, in many cases, be used to analyze a large number of experimental images in a relatively short time [27,28,29]. However, traditional object tracking methods proved to be inadequate for analyzing the evolution and motion of inclusions on smectic bubbles, motivating us to explore using machine learning techniques. We started, for simplicity, with the classification of circular inclusions on flat smectic films. By using Canny edge detection coupled with a support vector machine, we have developed a method that is capable of detecting features in these experimental systems more accurately and consistently than traditional techniques.

Successful object identification using TrackPy and its IDL progenitor (and most other tracking programs) typically requires careful tailoring of the tracking parameters to each new set of experimental images in order to accommodate variations in image quality and background texture, and spatial variations in brightness. When the target objects can easily be distinguished against a uniform background, conventional tracking programs generally perform well. Smectic films, however, often contain islands and regions of different thickness and consequently different brightness [5]. In addition, the films studied here include small, bright oil droplets. Even with extensive and time-consuming adjustment of the image processing and object identification parameters, conventional tracking programs are often unable to classify objects of interest with sufficient accuracy in complex images such as these.

Machine learning techniques have previously been shown to be effective at identifying and tracking objects in a variety of soft matter systems [30,31,32,33,34], including classifying textural features to determine the director orientation on the surface of a nematic liquid crystal [35] and detecting topological defects in the simulated textures of nematics confined to two dimensions [36]. In this paper, we describe a radial support vector machine (SVM), a widely used machine learning method for classification, regression, and other learning tasks [37], that we have developed for identifying and tracking large numbers of circular inclusions embedded in smectic A liquid crystal films. The SVM was found to achieve accurate classification of polydisperse emulsions of islands and oil droplets even when these inclusions were embedded in inhomogeneous background films.

## II. EXPERIMENTAL

Smectic liquid crystal films were drawn across an 8-mm hole in a microscope cover slip mounted in a temperature-controlled hot stage (Instec Inc. Model HCS302). We used the DisplayTech room-temperature smectic A mixture MX12160, which has the bulk phase sequence Isotropic 51.1°C SmA –3.2°C Crystal, doped with 0.9 – 1.9% paraffin oil by mass. The transition temperatures in the films were observed to be similar to the bulk values. The films were drawn at 30°C, where they were homogeneous in appearance. The films were allowed to equilibrate for a few minutes, and were then heated rapidly to 47°C, causing some of the paraffin oil to phase-separate from the liquid crystal and form a dense emulsion of droplets on the film. This heating protocol helped to ensure that the background film generally remained of uniform thickness while the droplets formed. With slower heating, the prepared films also formed islands (localized, thicker areas of the film bounded by layer dislocations) and layer steps. Images of islands were also obtained for training purposes in smectic A films of neat 8CB (4′-*n*-octyl-4′-cyanobiphenyl). The films were viewed in reflected light using a 20X magnification objective.

## III. SVM TRAINING AND INITIALIZATION

The support vector machine was developed using the Scikit-learn Python library [38], which provides many useful tools for the implementation of machine learning techniques. The SVM was trained using 1256 cropped images of MX12160 and 8CB films, with roughly half of these images containing either a single island or droplet, and the rest containing miscellaneous features typically seen in experiments (such as the edge of the microscope field stop and curved layer steps) that were designated as noise.

The SVM used these images to create and actively refine a separation boundary used to classify objects of interest based on their deviation in size or shape from an "ideal" droplet or island. The decision boundary that separates relevant and non-relevant features is determined by the "support vector" data points that define the maximum margin hyperplane. Once trained, the SVM was validated on a testing set of data in order to estimate the method's training accuracy, yielding a value of 96%. When the support vector machine analyzes new images not in the training set, it automatically categorizes each object as a positive or negative hit and records its decisions for future analyses. If an object is mis-categorized by the SVM, the exact feature that was misclassified can then be found and reviewed, and if necessary, added to the future training data to refine the decision boundary. This means that the SVM can adapt to new data very well, and multiple models can be trained to work on different types of images.

The experimental images were preprocessed using Canny edge detection to delineate the boundaries of potential objects of interest. For each such object, the number of vertices (pixels where none of the direct neighbors has similar intensity) in the boundary was calculated. A linear vertex density was computed for each such object by comparing the number of vertices to the total length of the curved edge of the object: objects with high vertex density are suggestive of circles, the canonical shape of islands and droplets, while objects with low density are characteristic of layer steps and the straight edges of the microscope field stop. Candidate curved objects of interest were then filtered according to their total length and enclosed area in order to eliminate extremely small or large objects from further consideration. Sub-images containing the remaining potential objects of interest were then extracted and passed to the support vector machine, which decides, based on the object's resemblance to objects in the training data, whether this is a feature of interest or should be discarded. By performing these operations over successive video frames, it is possible to track the changes in the size and position of these objects over time. The performance of the SVM was judged by whether it could accurately identify and track both islands and droplets on smectic films with significant spatial variations in brightness.

## IV. RESULTS

The accuracies of TrackPy and the SVM in identifying islands and oil drops were evaluated for a large number of images of smectic films by comparing the number of features detected by the software to the number of features identified by eye. Five representative images of MX12160 films, chosen to include typical features that make object detection challenging using traditional methods, are shown in Figs. 1 to 3.

In a typical application of TrackPy, the experimental images are first converted to monochrome and then thresholded to an appropriate level, yielding binary images in which the features to be tracked are clearly visible. Features of interest are then identified based on their brightness and area. In our comparisons, the TrackPy parameters were carefully chosen in order to compensate for whether the features of interest in each image were brighter or darker than the surrounding film, to optimize their contrast with the background, and to limit the size range of the objects to be considered.

The SVM method was found to be demonstrably superior in all of these examples (see Table 1), detecting droplets and islands significantly more accurately than TrackPy. The only cases in which the two methods showed comparable performance was when the background film was essentially uniform and the objects of interest were visually distinct, such as in the image shown in Fig. 1, where well dispersed oil droplets are clearly visible on a green background. In the other test cases, the support vector machine was found to be vastly superior. The SVM was reliably able to identify individual droplets in close-packed chains trapped at layer steps, to accommodate a range of different colors and intensities occurring within a single image, and to identify nested islands and droplets, as shown in Figs. 2 and 3. Although more computationally intensive, the support vector machine was far more accurate at identifying features of interest in inhomogeneous films than TrackPy.

## V. DROPLET DYNAMICS: CONVECTION OF AN OIL DROP EMULSION

To illustrate the utility of the SVM tracking method in tracking inclusion motion, an eight-second video of paraffin oil droplets exhibiting convective motion on a smectic A film was analyzed using the SVM. The locations of the inclusions in the first frame in this series (image #2) are shown in Fig. 2. The size and position of the droplets were determined for all of the frames in the image sequence and the droplet locations exported to TrackPy to determine their motion. As is evident from the trajectories shown in Fig. 4, we were able to identify and track all of the small oil droplets accurately throughout the entire image sequence. The SVM did not consistently identify the two larger droplets because of the bright, extended reflections visible near the tops of these droplets, which have non-uniform intensity and varying contrast around their edges. As a result, these regions do not appear to be circular and are discarded by the SVM following Canny edge detection analysis.

## VI. SUMMARY

We have developed a radial support vector machine that enables the reliable identification of circular inclusions in inhomogeneous smectic liquid crystal films, giving a roughly two-fold improvement in overall accuracy over TrackPy for all but the simplest experimental images. The SVM was found to be markedly superior at identifying target features on films of non-uniform appearance (such as those containing layers steps or with intensity gradients), and at detecting objects of significantly different brightness, nested objects (such as droplets within islands), and features of interest that are in close contact (such as dense droplet clusters and chains). The fully trained SVM eliminates the need for custom image processing for different sets of experimental images and is largely insensitive to intensity gradients and other visual artefacts. The SVM approach to identifying islands and droplets described here is extremely versatile and could be applied, with appropriate training images, to identifying a range of features of interest embedded in complex background textures in a variety of soft and biological systems.

## ACKNOWLEDGEMENTS

This work was supported by NASA Grants NAGNNX07AE48G and NNX-13AQ81G, and by the Soft Materials Research Center under NSF MRSEC Grants DMR-0820579 and DMR-1420736.

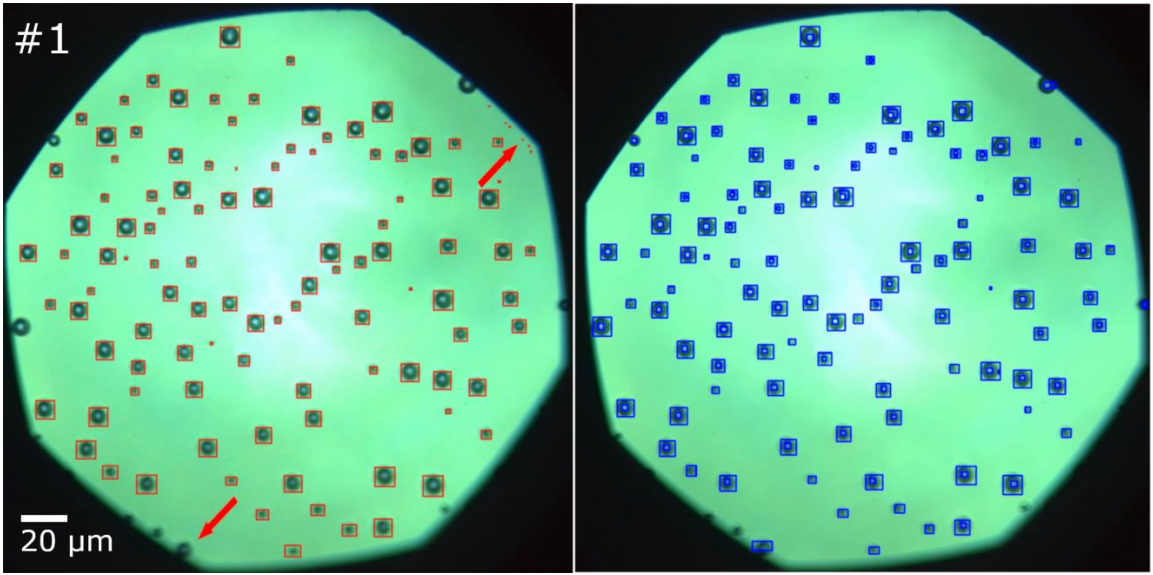

**Fig. 1.** Detection of oil droplets on a thin, homogeneous smectic A film viewed in reflection, using TrackPy (red boxes) and the SVM (blue boxes). The bounding boxes show positive detections. TrackPy performs well on this kind of image but misses droplets near the lower left edge of the image (red arrow at the lower left) and gives several false positives along the upper right edge (red arrow at the upper right). The SVM identifies virtually all of the droplets, with no false positives. The field of view is limited here and in Fig. 2 by the field stop of the microscope. The bright flare in the middle of the image is an illumination artifact.

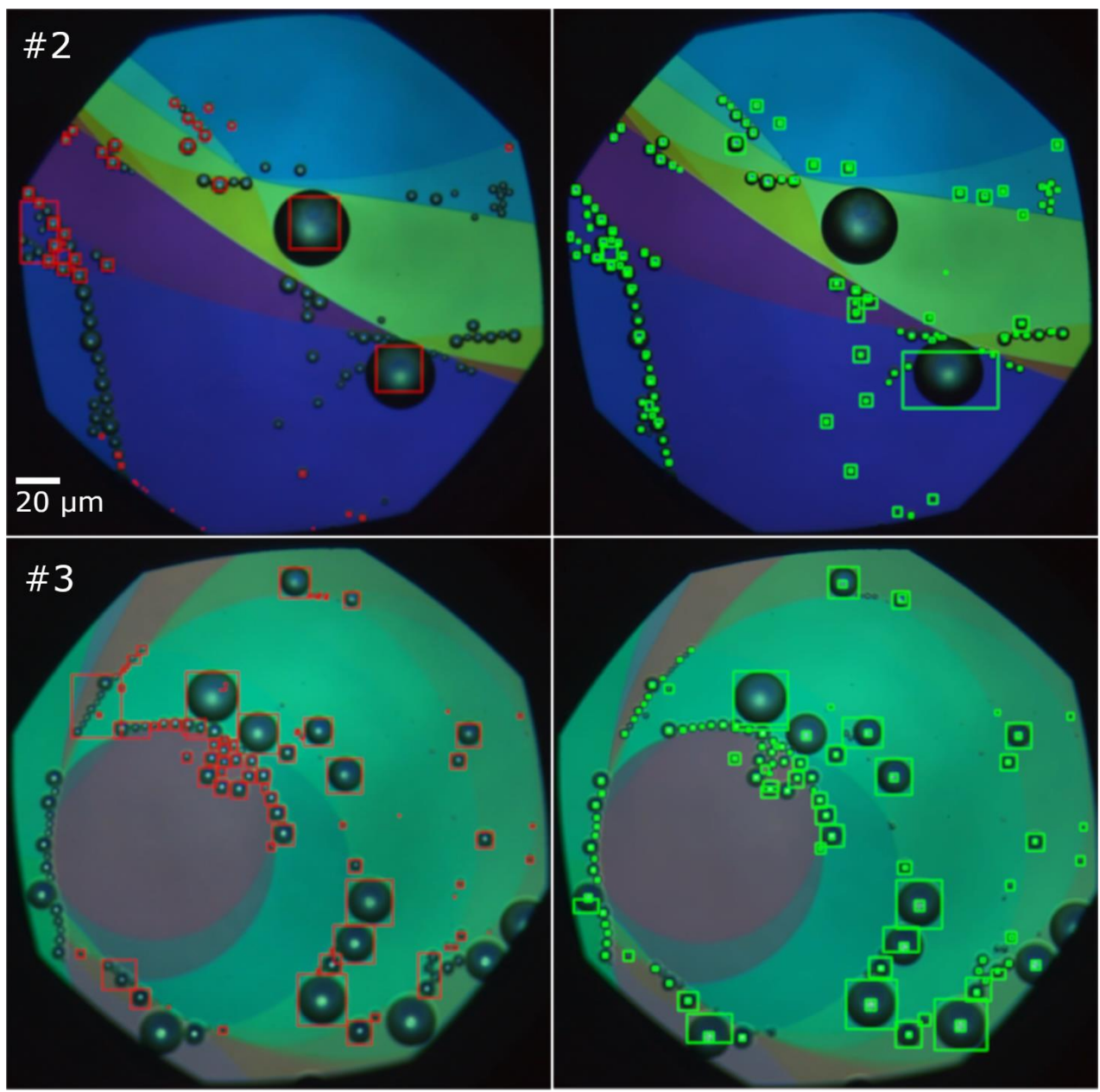

**Fig. 2.** Detection of oil droplets on two inhomogeneous smectic A films viewed in reflection, using TrackPy (red boxes) and the SVM (green boxes). Regions with different thickness bounded by dislocations in the smectic layering (layer steps) have different interference colors. Some oil droplets are observed in homogeneous regions of the film but most are located along layer steps, where they typically form compact chains. The SVM is far more successful than TrackPy in identifying the droplet locations correctly in both examples, with the exception of the two large droplets in film #2. TrackPy is unable to identify individual oil droplets in chains near the lower left edges of both images, and even fails to detect some droplets that are in homogeneous regions of the film.

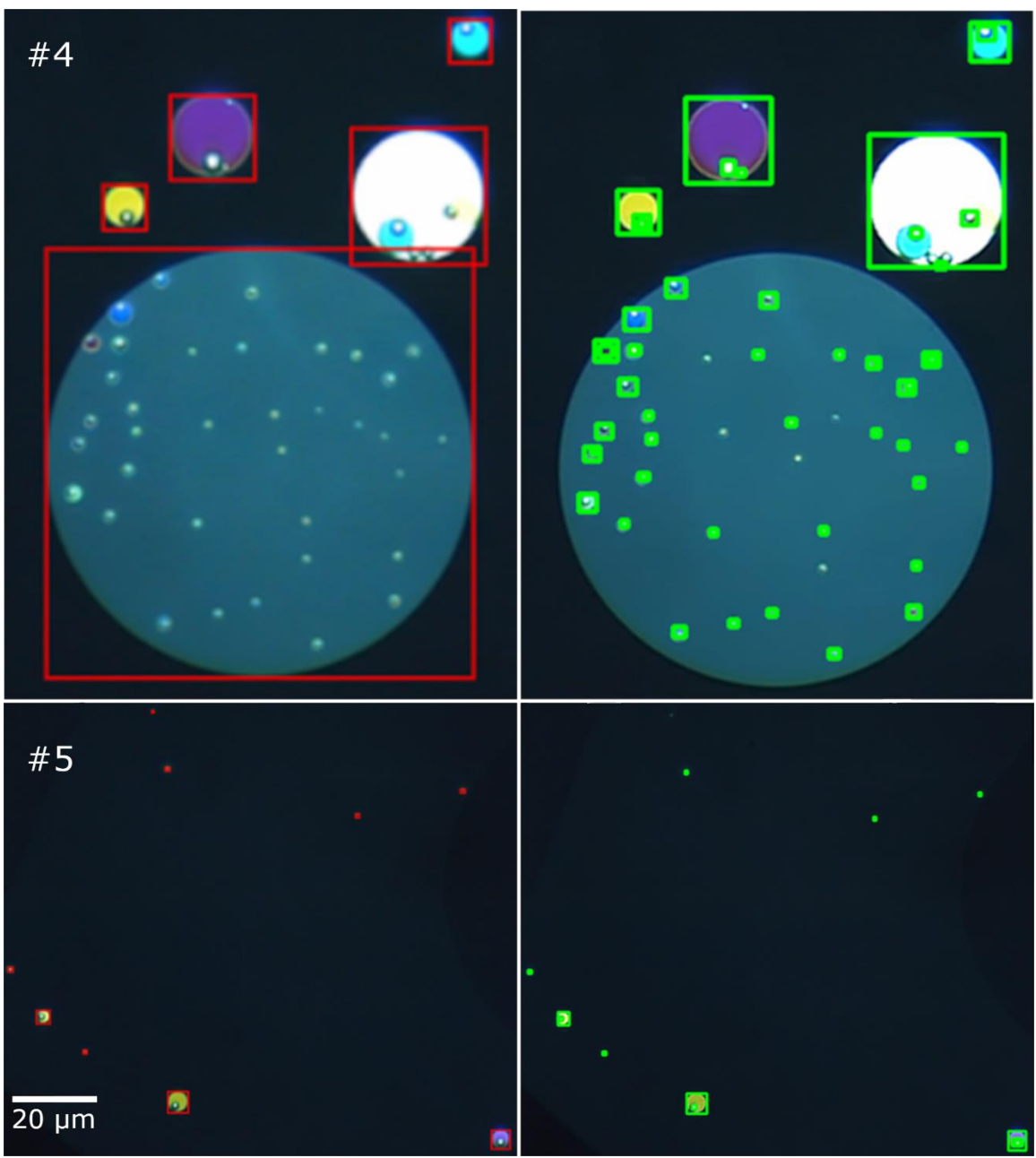

**Fig. 3.** Detection of oil droplets and islands on thin, homogeneous smectic A films viewed in reflection, using TrackPy (red boxes) and the SVM (green boxes). The smaller, inhomogeneous dots are droplets and the uniformly colored disks are islands. TrackPy successfully detects all of the islands in both images, while the SVM misses the biggest island in the first image and one tiny island near the upper edge of the second image. This could be remedied by manually moving the images of these particular islands to the "Hits" folder and retraining the machine. In the first image, TrackPy fails to identify any of the oil droplets that are contained within islands, features that are identified consistently by the SVM.

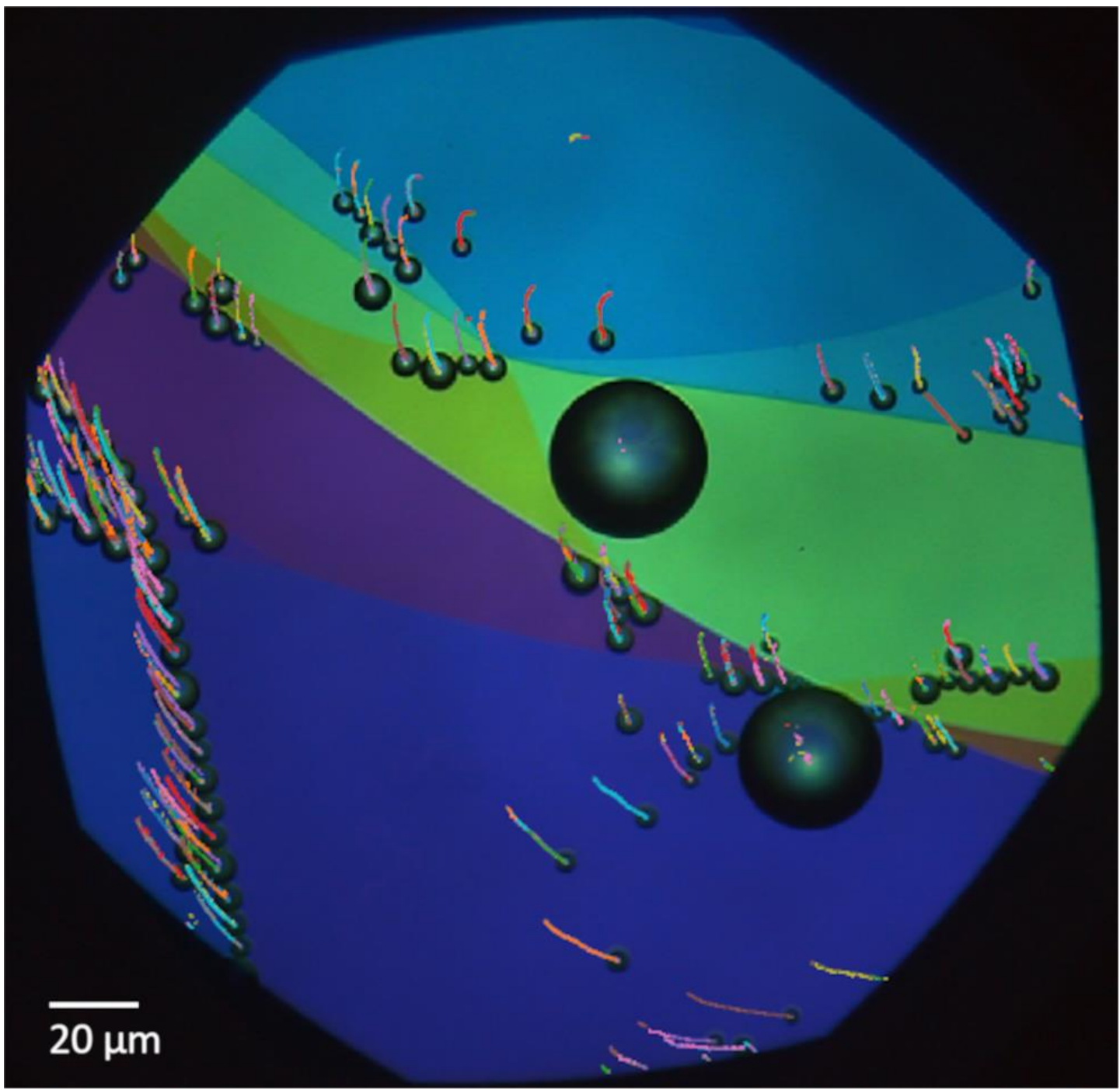

**Fig. 4.** Trajectories of oil droplets moving on a smectic A film viewed in reflection. A series of 302 consecutive video frames (corresponding to an elapsed time of eight seconds) were passed through the support vector machine to identify the locations of the oil droplets. The droplet trajectories were then computed by linking their locations in TrackPy.

| Image # | Accuracy (%) | | Runtime (seconds) | |
| --- | --- | --- | --- | --- |
| | TrackPy | SVM | TrackPy | SVM |
| 1 | 91.7 | 95.4 | 4.1 | 22.4 |
| 2 | 23.2 | 100 | 4.3 | 21.9 |
| 3 | 37.6 | 83.2 | 4.3 | 21.9 |
| 4 | 7.8 | 90.2 | 4.0 | 16.6 |
| 5 | 75.0 | 91.6 | 4.1 | 14.9 |

**Table 1.** Comparison of TrackPy and the SVM for detecting oil droplets and smectic islands in typical film images. The support vector machine achieves significantly higher accuracy but takes longer to analyze each image. Both programs were run on a Windows computer with a Ryzen 5 3600 AMD processor (6 core, 3.8 GHz), 16 GB of DDR4 RAM, and an NVidia GTX 1650 Super (4GB DDR6, 1880 MHz) graphics card.